\documentclass[aps,prd,eqsecnum,preprint,tightenlines,nofootinbib,showpacs]{revtex4-1}
\usepackage{graphicx,latexsym}

\def\bea{\begin{eqnarray}}
\def\eea{\end{eqnarray}}
\def\be{\begin{equation}}
\def\ee{\end{equation}}
\newcommand{\ub}[1]{\underline{#1}}
\newcommand{\ob}[1]{\overline{#1}}
\newcommand{\Pminus}{{\cal P}^-}
\newcommand{\Pfree}{\Pminus_0}
\newcommand{\Pint}{\Pminus_{\rm int}}

\begin{document}

\title{Convergence of the light-front coupled-cluster method
in quenched scalar Yukawa theory
}
\author{Austin Usselman}
\author{Sophia S. Chabysheva}
\author{John R. Hiller}
\affiliation{Department of Physics and Astronomy\\
University of Minnesota-Duluth \\
Duluth, Minnesota 55812}

\date{\today}

\begin{abstract}
We explore the convergence of the light-front coupled-cluster
(LFCC) method in the context of two-dimensional quenched scalar
Yukawa theory.  This theory is simple enough for higher-order
LFCC calculations to be relatively straightforward.  The quenching
is to maintain stability; the spectrum of the full theory with
pair creation and annihilation is unbounded from below.  The basic
interaction in the quenched theory is only emission and absorption
of a neutral scalar by the complex scalar.  The LFCC method builds
the eigenstate with one complex scalar and a cloud of neutrals
from a valence state that is just the complex scalar and the
action of an exponentiated operator that creates neutrals.  The
lowest order LFCC operator creates one; we add the next order,
a term that creates two.  At this order there is a direct contribution
to the wave function for two neutrals and one complex scalar and
additional contributions to all higher Fock wave functions from the
exponentiation.  Results for the lowest order and this new second-order
approximation are compared with those obtained with standard Fock-state
expansions.  The LFCC approach is found to allow representation
of the eigenstate with far fewer functions than the number of
wave functions required in a converged Fock-state expansion.

\end{abstract}


\maketitle

\section{Introduction} \label{sec:Introduction}

The calculation of the bound states for a given quantum field theory is an
inherently nonperturbative problem.  Various methods can be applied, the 
best known being, of course, lattice (gauge) theory~\cite{lattice}.  Here we consider
a method based on a Hamiltonian formulation in light-front coordinates~\cite{Dirac,LFreviews}.
The fundamental bound-state problem is then the eigenvalue problem 
\be
\Pminus|\psi(\ub{P})\rangle=\frac{M^2+P_\perp^2}{P^+}|\psi(\ub{P}\rangle,
\ee
where $\Pminus$ is the light-front Hamiltonian, $M$ is the mass of the 
eigenstate, and $\ub{P}=(P^+,\vec{P}_\perp)$ is the light-front 
momentum.\footnote{We define light-front coordinates~\protect\cite{Dirac}
and momenta as $x^\pm=t\pm z$, $\vec{x}_\perp=(x,y)$, 
$p^\pm=E\pm p_z$, $\vec{p}_\perp=(p_x,p_y)$.  The mass-shell condition 
for the total momentum is then $M^2=P^+P^--P_\perp^2$.}
The Hamiltonian is constructed from the Lagrangian ${\cal L}$ for a 
generic field $\phi$ as
\be
\Pminus=\int dx^-d^2x_\perp \left[:\frac{\delta{\cal L}}{\delta(\partial_+\phi)}-{\cal L}:\right]_{x^+=0}.
\ee
The eigenstate $|\psi(\ub{P})\rangle$ has definite momentum $\ub{P}$, and, once
known, can be used to compute properties of the state  This  formulation is 
particularly convenient for the computation of form factors, because $|\psi(\ub{P})\rangle$
is boost invariant.

The standard approach to the solution of the eigenvalue problem is to write
the eigenstate as a Fock-state expansion, which leads to a coupled system
of equations for the Fock wave functions.  This coupled system
is then converted into a matrix eigenvalue problem, either by
direct discretization, as in discrete light-cone quantization (DLCQ)~\cite{PauliBrodsky},
or by basis function expansion, as in basis light-front quantization (BLFQ)~\cite{Vary}.
However, a finite matrix representation requires a truncation of the Fock space.

This truncation has serious consequences.  In particular, there can be uncanceled
divergences, and self-energy corrections become dependent on the Fock sector and on
the presence of spectator constituents.  These are the nonperturbative analog of
what would happen to the contribution from a Feynman diagram if the diagram were
decomposed into the various time orderings, with the removal of the time 
orderings that involve too many intermediates.  These difficulties led to
the idea of sector-dependent renormalization~\cite{PerryWilson,HillerBrodsky,Karmanov},
which has its own difficulties~\cite{SecDep}.

As an alternative, we have developed the light-front coupled-cluster (LFCC) method~\cite{LFCC}.
No Fock-space truncation is invoked.  Instead, the eigenstate is written as coming from the
action of an exponentiated operator $T$ acting on a valence state $|\phi(\ub{P})\rangle$
\be
|\psi(\ub{P})\rangle=\sqrt{Z}e^T|\phi(\ub{P})\rangle,
\ee
with $\sqrt{Z}$ a normalization factor.\footnote{This construction was inspired by
the coupled-cluster method used in many-body problems of nuclear physics and 
quantum chemistry~\protect\cite{ManybodyCC}.}  The valence state is something simple that 
carries all the appropriate quantum numbers, in addition to the total momentum; for a
proton in QCD it would be the three-quark state.  The operator $T$ increases particle
number in various ways and conserves all the quantum numbers of the valence state; in QCD, 
$T$ would include gluon emission from a quark or gluon and pair creation from a gluon.  

The original eigenvalue problem is converted into
two parts, through multiplication by $e^{-T}$ and projection onto the valence sector
and its complement.  To express this, we define the effective Hamiltonian
$\ob{\Pminus}\equiv e^{-T}\Pminus e^T$ and the projection $P_v$ onto the valence
sector.  We then have
\be
P_v\ob{\Pminus}|\phi(\ub{P})\rangle=\frac{M^2+P_\perp^2}{P^+}|\phi(\ub{P})\rangle,\;\;
(1-P_v)\ob{\Pminus}|\phi(\ub{P})\rangle=0.
\ee
Roughly speaking, the first equation determines $M$ and any wave functions in 
$|\phi\rangle$, while the second determines the functions that define the structure of $T$.
In reality, of course, they are a coupled system, unless the valence state has
a single constituent and therefore no wave functions.  

All of this is obviously more complicated than the original eigenvalue problem,
but it is exact.  The power of the approach comes from the approximation step: Rather than
truncate Fock space, we truncate $T$.  Even for the simplest $T$ operator, its
exponentiation allows the eigenstate to span an infinite Fock space, and, without much
difficulty, one can arrange the approximate $e^T|\phi\rangle$ to fully explore all Fock sectors
relevant for the quantum numbers of the valence state.  In terms of
a Fock-state expansion, what we have done is to force the wave functions of
the higher Fock sectors to be directly dependent on those of the lower sectors
rather than setting these higher wave functions to zero, as would happen in a Fock-space
truncation.  Yet another way to interpret the LFCC approximation is that the
eigenstate is represented by a generalized coherent state.  In any case, the avoidance
of a Fock-space truncation eliminates the sector dependence and spectator
dependence of self-energy corrections and potentially controls the uncanceled
divergences.

The LFCC equations themselves are also truncated.  The complement projection $1-P_v$ is
restricted to the lowest set of Fock sectors necessary to have enough equations
to solve for the functions that define $T$.  This means that the LFCC method is 
not variational; the effective Hamiltonian $\ob{\Pminus}$ is not Hermitian, and the
truncated projections are not equivalent to minimization of the
expectation value $\langle \psi|\Pminus |\psi\rangle$.

One price to be paid for the gains of the LFCC method is that the LFCC equations are 
nonlinear.  The existence of a solution can be difficult to guarantee.  However, a 
linearized perturbative solution shows that the LFCC equations re-sum perturbation theory
to all orders for a restricted set of diagrams.  (The restriction arises because of
the truncation of $T$.)  This implies that, for weak coupling, a physical solution 
must exist.  Depending on the structure chosen for $T$ and $|\phi\rangle$, the physical 
solution may disappear as the coupling is increased.  An explicit example of this appears 
in an application to $\phi^4$ theory~\cite{LFCCphi4}, where the solution for the 
lowest-order approximation for $T$ does not extend beyond a certain coupling 
strength.  This is likely due to the restriction of the valence state to a
single constituent in a regime near the critical coupling where all Fock
sectors should contribute strongly.

One question that immediately arises has to do with the convergence of
the method, in the sense that as one relaxes the truncations of $T$ and $1-P_v$,
how does the solution improve?  The present work answers this question in
a particular context, with an application to quenched scalar Yukawa theory~\cite{WC}
in two dimensions.\footnote{The restriction to two dimensions is to disentangle
the convergence question from regularization and renormalization issues.  The quenching,
to eliminate pair production, is necessary for the theory to have a spectrum bounded
from below~\protect\cite{Baym}.}  In general, the correspondence between perturbation
theory and the LFCC resummation at weak coupling shows that the convergence
of the LFCC method is closely related to the convergence of perturbation
theory at weak coupling.  To get beyond weak coupling, we compare a
nonperturbative Fock-state expansion calculation to LFCC calculations
done with $T$ operators of increasing complexity.

The quenching of the theory eliminates potential concerns about the
vacuum.  Recent work~\cite{BCH,Collins,Martinovic,ETtoLF,Katz}
has emphasized the need for care in considering
the vacuum on the light front, but here no vacuum bubbles can occur.  
This also means that the Fock wave functions of a massive state do 
not include vacuum contributions and therefore have a direct physical 
interpretation. This is not generally true in equal-time quantization, where one
must compute the vacuum state as well as massive states; as an example, 
see the work on $\phi^4$ theory by Rychkov and Vitale~\cite{RychkovVitale}.

The Lagrangian, Hamiltonian, and Fock-state expansion for quenched
scalar Yukawa theory are given in Sec.~\ref{sec:qsy}.  The formulation
of the LFCC method for this theory is developed in Sec.~\ref{sec:LFCC}.
The results for both the Fock-state expansion method and the LFCC method
are presented and compared in Sec.~\ref{sec:results}, with
a brief summary provided in Sec.~\ref{sec:summary}.  Details
of numerical methods and diagrammatic rules are left to
appendices.

\section{Quenched scalar Yukawa theory} \label{sec:qsy}

The Lagrangian for scalar Yukawa theory~\cite{WC} is
\be
{\cal L}=|\partial_\mu\chi|^2-m^2|\chi|^2+\frac12(\partial_\mu\phi)^2-\frac12\mu^2\phi^2-g\phi|\chi|^2,
\ee
where $\chi$ is a complex scalar field with mass $m$ and $\phi$ is a real scalar field
with mass $\mu$.  The two fields are coupled by a Yukawa term with strength $g$.
In two dimensions, the light-front Hamiltonian density is
\be
{\cal H}=m^2|\chi|^2+\frac12\mu^2\phi^2+g\phi|\chi|^2.
\ee
The mode expansions for the fields are\footnote{Beginning here and for the remainder 
of the paper the $+$ superscript of the light-front momentum is suppressed.}
\bea
\chi&=&\int \frac{dp}{\sqrt{4\pi p}}\left[c_+(p)e^{-ipx^-/2}+c_-^\dagger(p)e^{ipx^-/2}\right], \\
\phi&=&\int \frac{dp}{\sqrt{4\pi p}}\left[a(p)e^{-ipx^-/2}+a^\dagger(p)e^{ipx^-/2}\right].
\eea
The nonzero commutation relations of the creation and annihilation operators are
\be
[c_\pm(p),c^\dagger_\pm(p')]=\delta(p-p'), \;\;
[a(p),a^\dagger(p')]=\delta(p-p').
\ee
In terms of these operators, the quenched light-front Hamiltonian
$\Pminus=\int dx^- {\cal H}=\Pfree+\Pint$ is specified by
\be
\Pfree=\int dp \frac{m^2}{p}\left[c_+^\dagger(p)c_+(p)+c_-^\dagger(p)c_-(p)+a^\dagger(p)a(p)\right]
\ee
and
\be
\Pint=g\int \frac{dp dq}{\sqrt{4\pi p q (p+q)}}
  \left\{ \left[c_+^\dagger(p+q)c_+(p)+c_-^\dagger(p+q)c_-(p)\right]a(q) +{\rm h.c.}\right\}.
\ee
Pair creation and annihilation terms are suppressed, to stabilize the spectrum.

We seek eigenstates of $\Pminus$, for which the two-dimensional light-front mass
eigenvalue problem is
\be
\Pminus|\psi(P)\rangle=\frac{M^2}{P}|\psi(P)\rangle.
\ee
We limit this to the charge-one sector.
In the next section, we consider the LFCC approach to the solution of
this eigenvalue problem,
but here we develop the standard Fock-state expansion approach,
to use as a basis for comparison.  

We write the Fock-state expansion of the eigenstate as
\be  \label{eq:FSexpansion}
|\psi(P)\rangle=\sum_{n=0}^\infty P^{n/2}\int\left(\prod_{i=1}^n dx_i\right)
  \theta(1-\sum_i x_i)\psi_n(x_1,\ldots,x_n)\frac{1}{\sqrt{n!}}
     \prod_i a^\dagger(x_iP)c_+^\dagger((1-\sum_i x_i)P)|0\rangle.
\ee
Projection of the eigenvalue problem onto 
$ \prod_j^{n'} a^\dagger(y_jP)c_+^\dagger((1-\sum_j^{n'} y_i)P)|0\rangle$,
and division by $\mu^2$,
yields coupled equations for the Fock-state wave functions $\psi_n$
\bea  \label{eq:FSsystem}
\lefteqn{\left[\frac{\tilde{m}^2}{1-\sum_j y_j}+\sum_j\frac{1}{y_j}\right]\psi_n(y_1,\ldots,y_n)
+\frac{\lambda}{\sqrt{n}}\sum_j^n\frac{\psi_{n-1}(y_1,\ldots,y_{j-1},y_{j+1},\ldots,y_n)}
                                       {\sqrt{y_j(1-\sum_{i\neq j}y_i)(1-\sum_i^n y_i)}}}&& \\
&&+\lambda\sqrt{n+1}\int dx\, \theta(1-x-\sum_i y_i)\frac{\psi_{n+1}(y_1,\ldots,y_n,x)}
                                                  {\sqrt{x(1-x-\sum_i y_i)(1-\sum_i y_i)}}
                                                  =\frac{M^2}{\mu^2}\psi_n(y_1,\ldots,y_n). \nonumber
\eea
Here $\tilde{m}\equiv m/\mu$ is a dimensionless relative mass and
$\lambda\equiv g/(\sqrt{4\pi}\mu^2)$ is a dimensionless coupling strength.
We solve this system numerically by first truncating the Fock space at
$n=n_{\rm max}$ neutrals and expanding the wave functions in a symmetrized
monomial basis.  The details are discussed in Appendix \ref{sec:methods}, and
the results in Sec.~\ref{sec:results}.

The structure of the eigenstate is studied by considering the relative
probabilities for Fock sectors with different numbers of neutrals.  These 
are formed as the ratio
\be \label{eq:RelProb}
R_n\equiv\frac{1}{\psi_0^2}\int \prod_i dx_i\,\theta(1-\sum_i x_i)|\psi_n|^2.
\ee
Results for these ratios are shown in Sec.~\ref{sec:results}.

\section{Light-front coupled-cluster method} \label{sec:LFCC}

The LFCC method constructs the charge-one eigenstate in the form
\be
|\psi\rangle=\sqrt{Z}e^Tc_+^\dagger|0\rangle,
\ee
where $c_+^\dagger|0\rangle$ is the single-particle valence state.  
The $T$ operator is expanded in a sequence $T=\sum_n T_n$, with
\be
T_n=\int \prod_i^n dx_i dp \, p^{n/2} t_n(x_1,\ldots,x_n)
     \prod_i^n a^\dagger(x_ip)c_+^\dagger((1-\sum_i^n x_i)p) c_+(p).
\ee
The factor $p^{n/2}$ is included to keep $T_n$
dimensionless; $p$ is the natural scale, being the momentum
flowing through the operator.

The action of $T_n$ is to increase the number of neutrals
by $n$, and the exponentiation of $T$ provides for generation
of all possible (quenched) charge-one Fock states, even if
$T$ is truncated to only $T_1$.  Without truncation, the
functions $t_n$ provide for an exact solution, with a
duality between the $t_n$ and the Fock-state wave functions $\psi_n$.
However, without truncation the eigenvalue problem is
equivalent to an infinite coupled system of nonlinear equations for
these $t_n$.  

We can then study the convergence of the LFCC method as
the number of terms in $T$ is increased.
Here we consider the first two terms, $T_1$ and $T_2$, and 
compare results with those from the truncated Fock-state expansion.

\begin{figure}[ht]
\vspace{0.2in}
\centerline{\includegraphics[width=8cm]{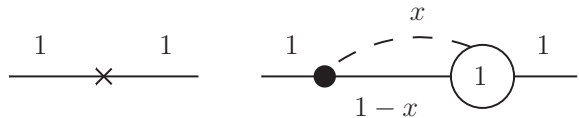}}
\caption{\label{fig:valence}
Diagrammatic representation of the valence equation.
Rules for diagrams are given in Appendix~\ref{sec:rules}.
}
\end{figure}

The LFCC form of the eigenvalue problem is
\be
P_v\ob{\Pminus} c_+^\dagger(P)|0\rangle=\frac{M^2}{P}c_+^\dagger(P)|0\rangle, \;\;
(1-P_v)\bar{\cal P}^- c_+^\dagger(P)|0\rangle=0.
\ee
Independent of the level of truncation for $T$, the first equation becomes
\be
\frac{m^2}{P}c_+^\dagger(P)|0\rangle 
+ \frac{g}{\sqrt{4\pi}}\int \frac{dq}{\sqrt{(P-q)qP}}
                    \frac{t_1(q/P)}{\sqrt{P}} c_+^\dagger(P)|0\rangle
= \frac{M^2}{P}c_+^\dagger(P)|0\rangle.
\ee
The contributions to this equation come from the $\Pfree$ and $\Pint T_1$
terms in $\ob{\Pminus}$, as represented diagrammatically in Fig.~\ref{fig:valence}.
On division by $\mu^2$, the projected valence equation reduces to
the following expression for the eigenmass $M$:
\be  \label{eq:M}
\frac{M^2}{\mu^2}=\tilde{m}^2+\lambda\int \frac{dx \,t_1(x)}{\sqrt{x(1-x)}}\equiv \tilde{m}^2+\lambda\Delta.
\ee
The self-energy term is then specified by
\be
\Delta=\int \frac{dx \,t_1(x)}{\sqrt{x(1-x)}}.
\ee
The function $t_1$ is to be obtained by solving the remaining LFCC
equations.

\begin{figure}[ht]
\vspace{0.2in}
\centerline{\includegraphics[width=12cm]{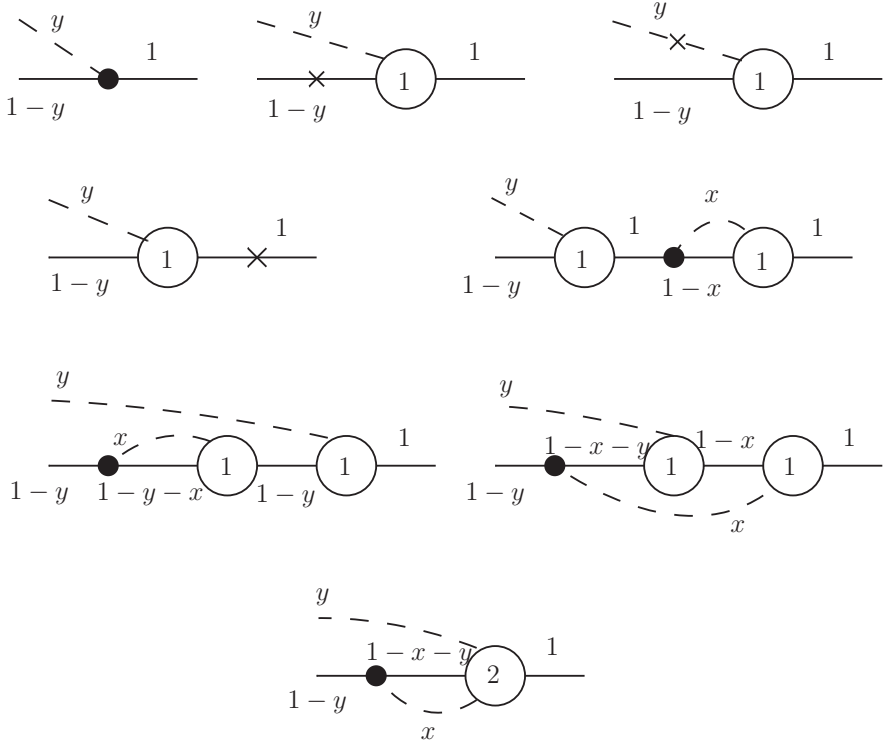}}
\caption{\label{fig:oneneutral}
Diagrammatic representation of the projection onto the
one-neutral Fock sector.
}
\end{figure}

With each truncation of $T$ there is a matching truncation of
the projector $1-P_v$ to include only enough Fock sectors to
determine the unknown functions in the retained terms of $T$.
Given the truncation to $T=T_1+T_2$, the equations for $t_1$ and $t_2$
take the form of two projections, onto the sectors with one
and two neutrals.  The contributions to the first projection come from 
\be \label{eq:oneneutral}
\ob{\Pminus}\rightarrow \Pint+\Pfree T_1-T_1 \Pfree-T_1 \Pint T_1
    +\frac12\Pint T_1^2 +\Pint T_2.
\ee
These terms are represented in Fig.~\ref{fig:oneneutral}
and yield the following equation for $t_1$:
\bea  \label{eq:t1}
0&=&\frac{\lambda}{\sqrt{y(1-y)}}
   +\left[\frac{\tilde{m}^2}{1-y}+\frac{1}{y}-\tilde{m}^2\right]t_1(y)
   -\lambda t_1(y)\int_0^1\frac{dx\,t_1(x)}{\sqrt{x(1-x)}}  \\
&& +\frac12\frac{\lambda}{1-y}t_1(y)\int_0^{1-y}\frac{dx\, t_1(\frac{x}{1-y})}{\sqrt{x(1-y-x)}} 
+\frac12\frac{\lambda}{\sqrt{1-y}}\int_0^{1-y}\frac{dx\,t_1(\frac{y}{1-x})t_1(x)}{\sqrt{x(1-x)(1-y-x)}}
\nonumber \\
&&+\frac{2\lambda}{\sqrt{1-y}}\int_0^{1-y}\frac{dx\,t_2(y,x)}{\sqrt{x(1-y-x)}}. 
\nonumber
\eea
The equation (\ref{eq:t1}) for $t_1$ can be obtained either by explicitly carrying out the
contractions of annihilation and creation operators or by diagrammatic rules
listed in Appendix~\ref{sec:rules}.

The first term in the second line of Eq.~(\ref{eq:t1}) can be simplified by rescaling
the integration variable $x$ by $1-y$; this shows the integral to be equal to $\Delta$.
The same self-energy integral appears in the last term of the first line.  The terms
proportional to $\Delta$ can then be collected with the $\tilde{m}$ terms, to 
introduce $M^2$ with use of (\ref{eq:M})
\bea  \label{eq:t1final}
\left[\frac{M^2}{\mu^2}-\frac{M^2/\mu^2}{1-y}-\frac{1}{y}\right]t_1(y)&=&\frac{\lambda}{\sqrt{y(1-y)}}
      +\frac{2\lambda}{\sqrt{1-y}}\int_0^{1-y}\frac{dx\,t_2(y,x)}{\sqrt{x(1-y-x)}}  \\
&& +\frac12\frac{\lambda}{\sqrt{1-y}}
\left[\int_0^{1-y}\frac{dx\,t_1(\frac{y}{1-x})t_1(x)}{\sqrt{x(1-x)(1-y-x)}}
       -\frac{\Delta}{\sqrt{1-y}}t_1(y)\right].
\nonumber
\eea
For the truncation $T=T_1$, this equation, with the $t_2$ term removed, is all
that need be solved.

The appearance of the physical mass $M$ in the invariant-mass terms on the left 
of (\ref{eq:t1final}) is typical of the LFCC method, where self-energy corrections are 
independent of the Fock sector and independent of spectators.  This avoids the use of 
the sector-dependent bare masses that are frequently introduced in truncated 
Fock-state-expansion calculations~\cite{PerryWilson,HillerBrodsky,Karmanov,SecDep}, 
where self-energy corrections are sector and spectator dependent.

The contributions to the second projection, onto the two-neutral sector, come
from the following terms in $\ob{\Pminus}$:
\bea
\ob{\Pminus}&\rightarrow& \Pfree T_2-T_2\Pfree
+\frac12\Pfree T_1^2-T_1 \Pfree T_1+\frac12 T_1^2 \Pfree
+\Pint T_1-T_1 \Pint \\
&&+\frac16\Pint T_1^3-\frac12 T_1 \Pint T_1^2+\frac12 T_1^2 \Pint T_1
+\frac12\Pint T_1 T_2+\frac12\Pint T_2 T_1-T_1\Pint T_2 - T_2\Pint T_1. 
\nonumber
\eea
Graphical representations of these terms are given in Figs.~\ref{fig:P0T2}-\ref{fig:PintT2T1}.
They and the rules in Appendix~\ref{sec:rules} yield the equation for $t_2$ as
\bea  \label{eq:t2}
0&=&2\left[\frac{\tilde{m}^2}{1-y_1-y_2}+\frac{1}{y_1}+\frac{1}{y_2} 
              -\tilde{m}^2\right]t_2(y_1,y_2) \\
&&
+\frac12\frac{t_1(\frac{y_2}{1-y_1})t_1(y_1)}{y_1\sqrt{1-y_1}}
    +\frac12\frac{t_1(\frac{y_1}{1-y_2})t_1(y_2)}{y_2\sqrt{1-y_2}}
    +\frac12\frac{t_1(\frac{y_2}{1-y_1})t_1(y_1)}{y_2\sqrt{1-y_1}}
    +\frac12\frac{t_1(\frac{y_1}{1-y_2}t_1(y_2)}{y_1\sqrt{1-y_2}} 
\nonumber \\
&&
+\frac12\frac{\tilde{m}^2 t_1(\frac{y_2}{1-y_1})t_1(y_1)}{(1-y_1-y_2)\sqrt{1-y_1}}
    +\frac12\frac{\tilde{m}^2 t_1(\frac{y_1}{1-y_2})t_1(y_2)}{(1-y_1-y_2)\sqrt{1-y_2}} 
\nonumber \\
&& 
  -\frac{\tilde{m}^2t_1(\frac{y_2}{1-y_1})t_1(y_1)}{(1-y_1)^{3/2}}
  -\frac{\tilde{m}^2t_1(\frac{y_1}{1-y_2})t_1(y_2)}{(1-y_2)^{3/2}}
-\frac{t_1(\frac{y_2}{1-y_1})t_1(y_1)}{y_1\sqrt{1-y_1}}
    -\frac{t_1(\frac{y_1}{1-y_2})t_1(y_2)}{y_2\sqrt{1-y_2}}
\nonumber \\
&& 
+\frac12\frac{\tilde{m}^2t_1(\frac{y_2}{1-y_1})t_1(y_1)}{\sqrt{1-y_1}}
   +\frac12\frac{\tilde{m}^2t_1(\frac{y_1}{1-y_2})t_1(y_2)}{\sqrt{1-y_2}} 
\nonumber \\
&&
+\frac{\lambda t_1(y_1)}{\sqrt{y_2(1-y_1)(1-y_1-y_2)}}
    +\frac{\lambda t_1(y_2)}{\sqrt{y_1(1-y_2)(1-y_1-y_2)}}
\nonumber \\
&&
-\lambda\frac{t_1(\frac{y_2}{1-y_1})}{(1-y_1)\sqrt{y_1}}
  -\lambda\frac{t_1(\frac{y_1}{1-y_2})}{(1-y_2)\sqrt{y_2}}
\nonumber \\
&&
+\frac16\left[\int_0^{1-y_1-y_2}dx\frac{\lambda t_1(\frac{y_2}{1-y_1-x})t_1(\frac{y_1}{1-x})t_1(x)}
                                  {\sqrt{x(1-x)(1-y_1-y_2)(1-y_1-x)(1-y_1-y_2-x)}}  + (y_1\leftrightarrow y_2)\right]
\nonumber \\
&& 
+\frac16\left[\int_0^{1-y_1-y_2}dx\frac{\lambda t_1(\frac{y_2}{1-y_1-x})t_1(\frac{x}{1-y_1})t_1(y_1)}
                                  {\sqrt{x(1-y_1)(1-y_1-y_2)(1-y_1-x)(1-y_1-y_2-x)}}  + (y_1\leftrightarrow y_2)\right]
\nonumber \\
&& 
+\frac16\left[\int_0^{1-y_1-y_2}dx\frac{\lambda t_1(\frac{x}{1-y_1-y_2})t_1(\frac{y_2}{1-y_1})t_1(y_1)}
                                  {(1-y_1-y_2)\sqrt{x(1-y_1)(1-y_1-y_2-x)}}  + (y_1\leftrightarrow y_2)\right]
\nonumber \\
&& 
-\frac12\left[\int_0^{1-y_1}dx\frac{\lambda t_1(\frac{y_2}{1-y_1})t_1(\frac{x}{1-y_1})t_1(y_1)}
                                      {(1-y_1)^{3/2}\sqrt{x(1-y_1-x)}}   + (y_1\leftrightarrow y_2)\right]
\nonumber \\
&& 
-\frac12\left[\int_0^{1-y_1} dx \frac{\lambda t_1(\frac{y_2}{1-y_1})t_1(\frac{y_1}{1-x})t_1(x)}
                                        {(1-y_1)\sqrt{x(1-x)(1-y_1-x)}}  + (y_1\leftrightarrow y_2)\right]
\nonumber \\
&& 
+\frac12\left[\int_0^1 dx \frac{\lambda t_1(\frac{y_2}{1-y_1})t_1(y_1)t_1(x)}
                                  {\sqrt{x(1-x)(1-y_1)}}  + (y_1\leftrightarrow y_2)\right].
\nonumber \\
&&
+\int_0^{1-y_1-y_2}dx\frac{\lambda t_2(\frac{y_1}{1-x},\frac{y_2}{1-x})t_1(x)}
                           {(1-x)\sqrt{x(1-y_1-y_2)(1-y_1-y_2-x)}}
\nonumber \\
&&
+\left[\int_0^{1-y_1-y_2}dx\frac{\lambda t_2(\frac{x}{1-y_1},\frac{y_2}{1-y_1})t_1(y_1)}
                           {(1-y_1)\sqrt{x(1-y_1-y_2)(1-y_1-y_2-x)}}  + (y_1\leftrightarrow y_2)\right]
\nonumber \\
&&
+\int_0^{1-y_1-y_2}dx\frac{\lambda t_1(\frac{x}{1-y_1-y_2})t_2(y_1,y_2)}{(1-y_1-y_2)\sqrt{x(1-y_1-y_2-x)}} 
\nonumber \\
&&
+\left[\int_0^{1-y_1-y_2}dx\frac{\lambda t_1(\frac{y_2}{1-y_1-x})t_2(y_1,x)}
                               {\sqrt{x(1-y_1-x)(1-y_1-y_2)(1-y_1-y_2-x)}}  + (y_1\leftrightarrow y_2)\right]
\nonumber \\
&&
-2\left[\int_0^{1-y_1}dx\frac{\lambda t_1(\frac{y_2}{1-y_1})t_2(y_1,x)}{(1-y_1)\sqrt{x(1-y_1-x)}}  
                                + (y_1\leftrightarrow y_2)\right]
-2 \int_0^1 dx \frac{\lambda t_2(y_1,y_2)t_1(x)}{\sqrt{x(1-x)}}.
\nonumber
\eea
We solve these equations numerically, as discussed in Appendix~\ref{sec:methods}, both for $t_1$ alone
and the coupled system, for $t_1$ and $t_2$.

\begin{figure}[ht]
\vspace{0.2in}
\centerline{\includegraphics[width=9cm]{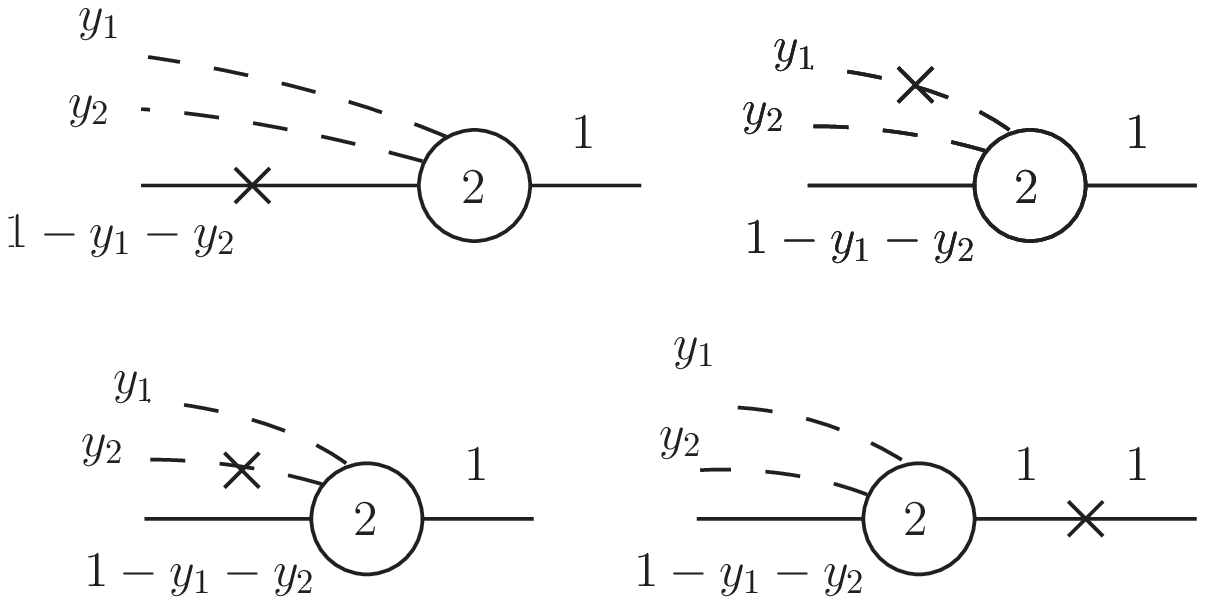}}
\caption{\label{fig:P0T2}
Diagrammatic representation of the projection onto the
two-neutral Fock sector of the $\ob{\Pminus}$ terms $\Pfree T_2-T_2\Pfree$.
}
\end{figure}
\begin{figure}[ht]
\vspace{0.2in}
\centerline{\includegraphics[width=15cm]{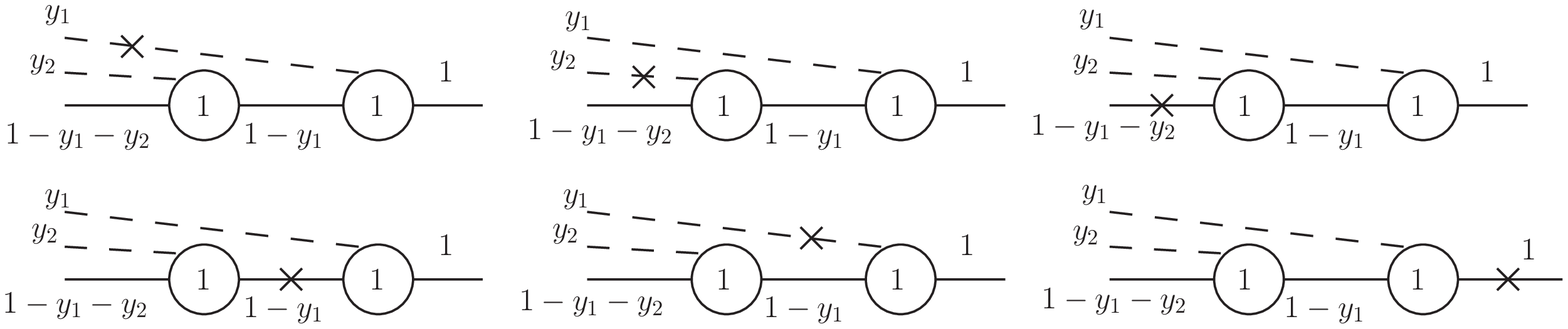}}
\caption{\label{fig:P0T12}
Same as Fig.~\ref{fig:P0T2} but for the $\ob{\Pminus}$ terms
$\frac12\Pfree T_1^2-T_1 \Pfree T_1+\frac12 T_1^2 \Pfree$.
}
\end{figure}
\begin{figure}[ht]
\vspace{0.2in}
\centerline{\includegraphics[width=10cm]{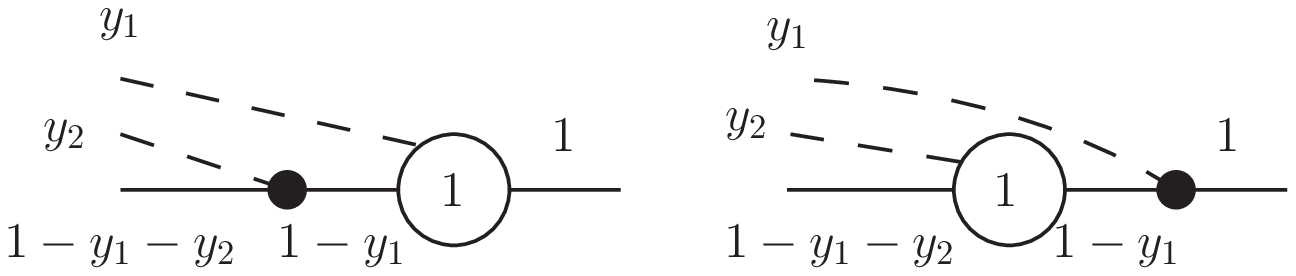}}
\caption{\label{fig:PintT1}
Same as Fig.~\ref{fig:P0T2} but for the $\ob{\Pminus}$ terms
$\Pint T_1-T_1 \Pint$.}
\end{figure}
\begin{figure}[ht]
\vspace{0.2in}
\centerline{\includegraphics[width=15cm]{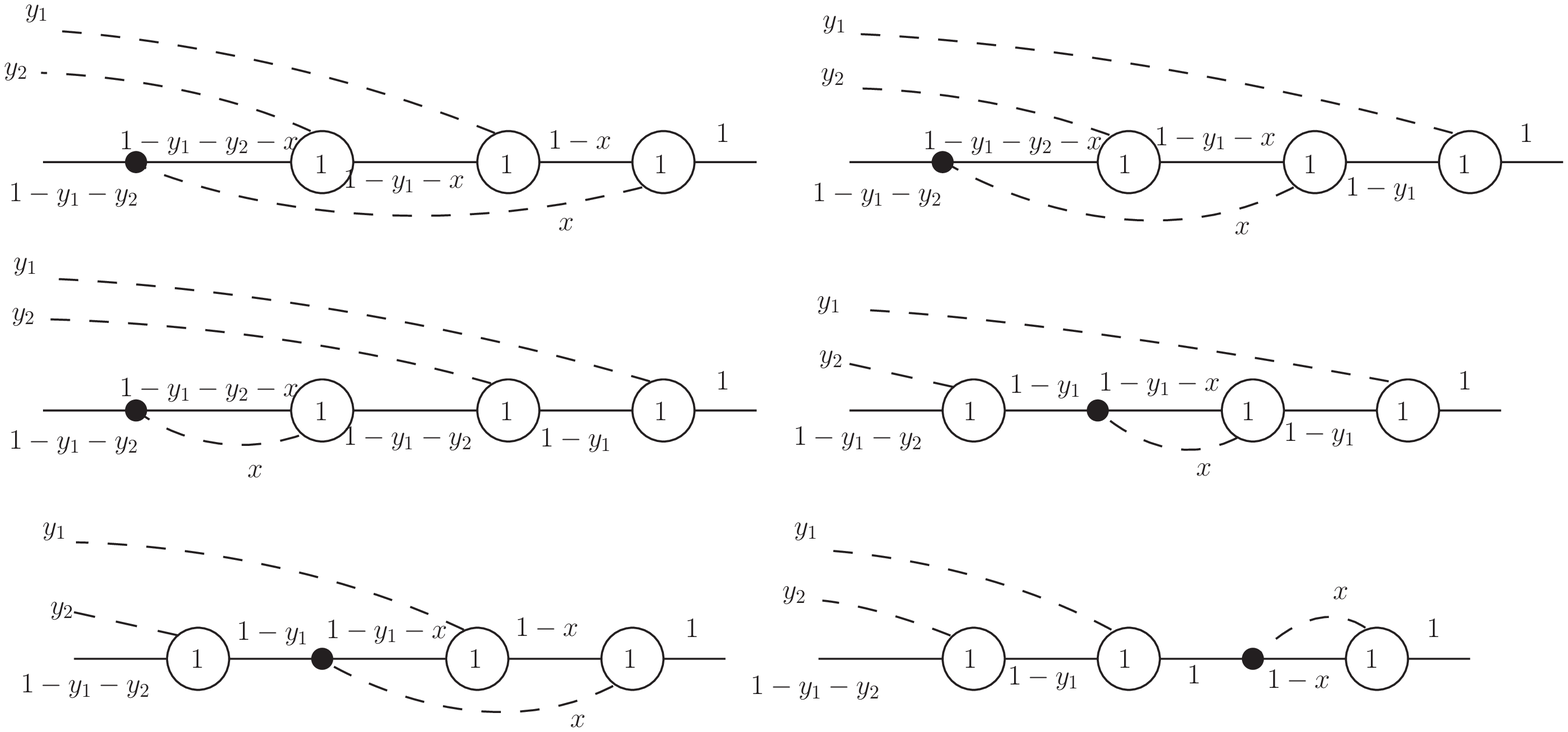}}
\caption{\label{fig:PintT13}
Same as Fig.~\ref{fig:P0T2} but for the $\ob{\Pminus}$ terms
$\frac16\Pint T_1^3-\frac12 T_1 \Pint T_1^2+\frac12 T_1^2 \Pint T_1$.
}
\end{figure}
\begin{figure}[ht]
\vspace{0.2in}
\centerline{\includegraphics[width=15cm]{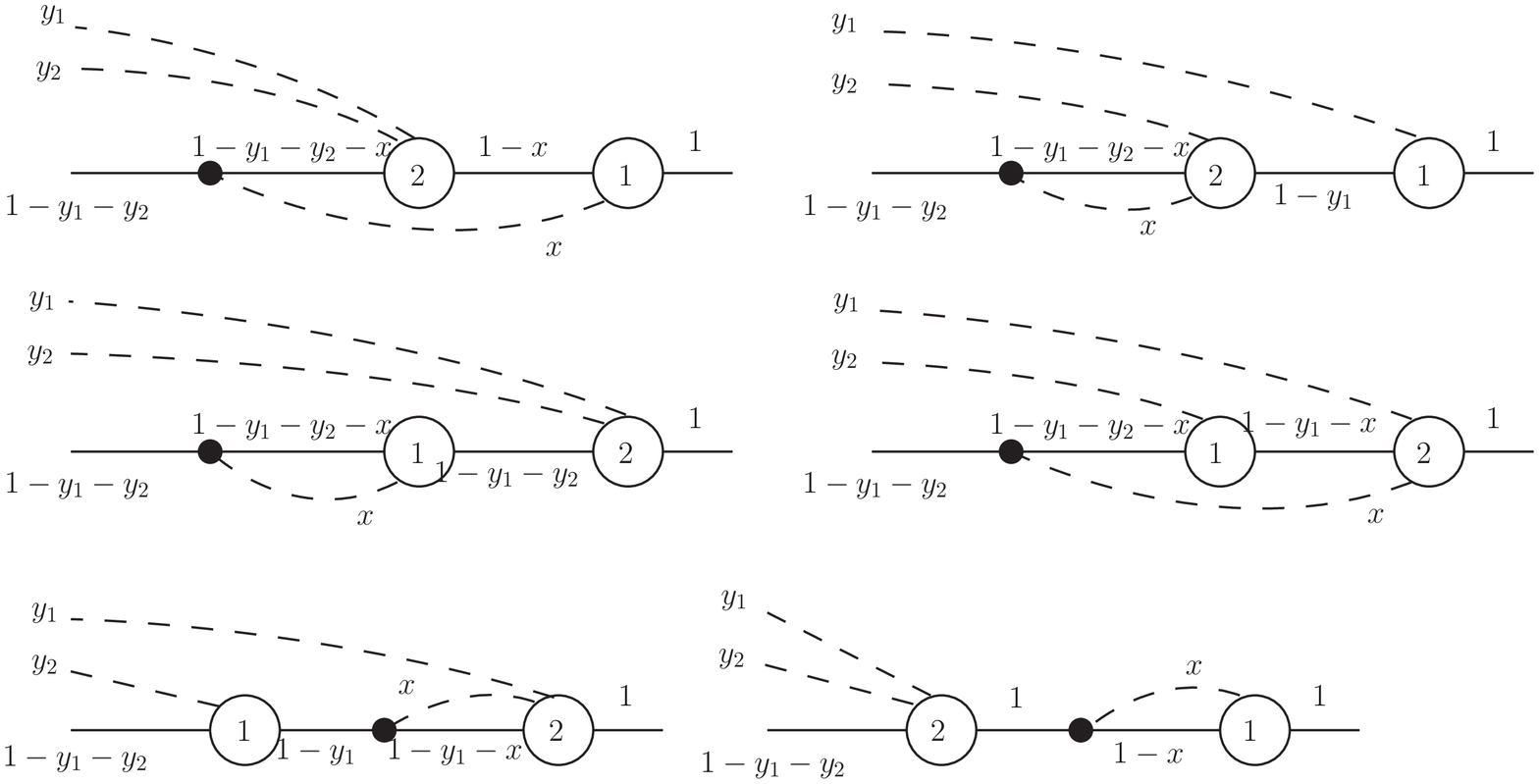}}
\caption{\label{fig:PintT2T1}
Same as Fig.~\ref{fig:P0T2} but for the $\ob{\Pminus}$ terms
$\frac12\Pint T_1 T_2+\frac12\Pint T_2 T_1-T_1\Pint T_2 - T_2\Pint T_1$.}
\end{figure}

The relative probabilities for different Fock sectors can be computed from the expansion of
the exponential form of the LFCC approximation
\be
|\psi\rangle=\sqrt{Z}e^{T_1+T_2}c_+^\dagger(P)|0\rangle
    \simeq\sqrt{Z}\left[1+T_1+(T_2+\frac12 T_1^2)+\cdots\right]c_+^\dagger(P)|0\rangle.
\ee
The Fock state wave functions can be extracted by comparison with the Fock state
expansion in (\ref{eq:FSexpansion}), after the actions of the operators $T_1$ and $T_2$ are
taken into account.  We find $\psi_0=\sqrt{Z}$, $\psi_1(x)=\sqrt{Z}t_1(x)$, and
\be
\psi_2(x_1,x_2)=\sqrt{\frac{Z}{2}}\left[2t_2(x_1,x_2)
                      +\frac{t_1(\frac{x_2}{1-x_1})t_1(x_1)}{\sqrt{1-x_1}}
                      +\frac{t_1(\frac{x_1}{1-x_2})t_1(x_2)}{\sqrt{1-x_2}}\right].
\ee
The relative probabilities for the one and two-neutral sectors can then be
computed as before, using (\ref{eq:RelProb}).  The necessary integrals can be
done analytically for the basis function expansions introduced in Appendix~\ref{sec:methods};
however, for the cross term between the second and third terms of $\psi_2$,
the analytic result is the value of a hypergeometric function and that term 
is instead integrated numerically with Gauss-Legendre quadrature.
The overall normalization $Z$ is not computable in a finite sum, which is the
motivation for considering relative probabilities, rather than absolutes.
Fock sectors higher than the two-neutral sector can be considered, but the
wave functions become much more complicated.

\section{Results} \label{sec:results}

\begin{figure}[ht]
\vspace{0.2in}
\centerline{\includegraphics[width=15cm]{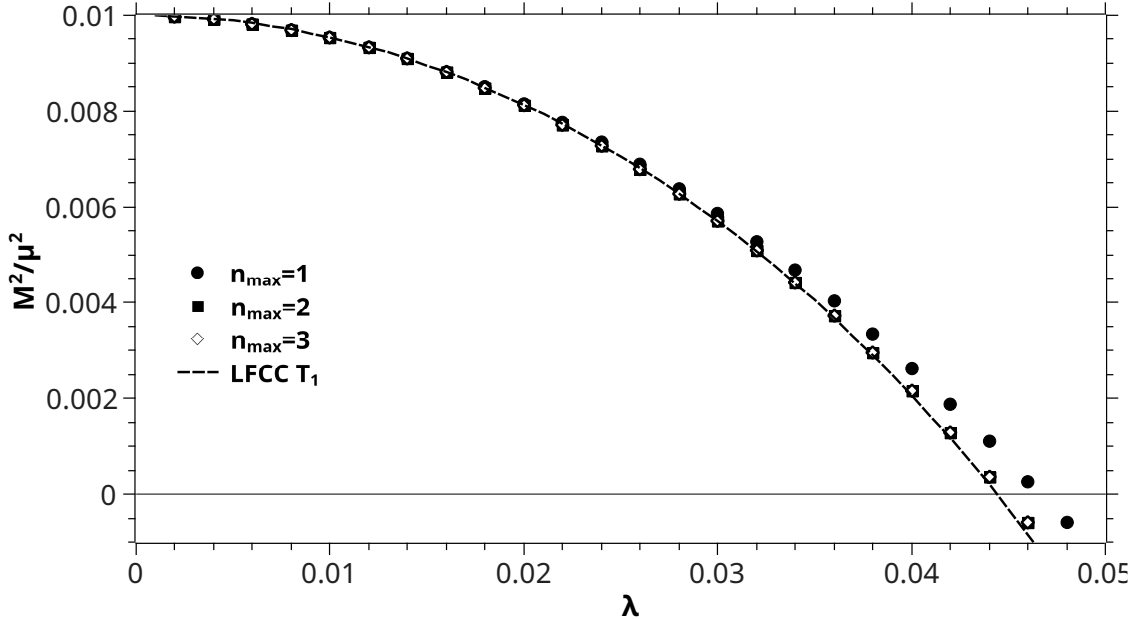}}
\caption{\label{fig:t1mt01}
The mass eigenvalue ratio $M^2/\mu^2$ as a function of the dimensionless
coupling $\lambda$ for a series of Fock-space truncations and for
the LFCC approximation $T=T_1$.  The mass
ratio of the constituents is $\tilde{m}\equiv m/\mu=0.1$.  The basis sets
in each Fock sector were limited to orders $N=10$, 14, and 7 for $n_{\rm max}=1$,
2, and 3, respectively.  The basis
set for the LFCC result has a maximum order of $N_1=9$.  Addition of the
$T_2$ operator does not significantly change the LFCC results.
}
\end{figure}

\begin{figure}[ht]
\vspace{0.2in}
\centerline{\includegraphics[width=15cm]{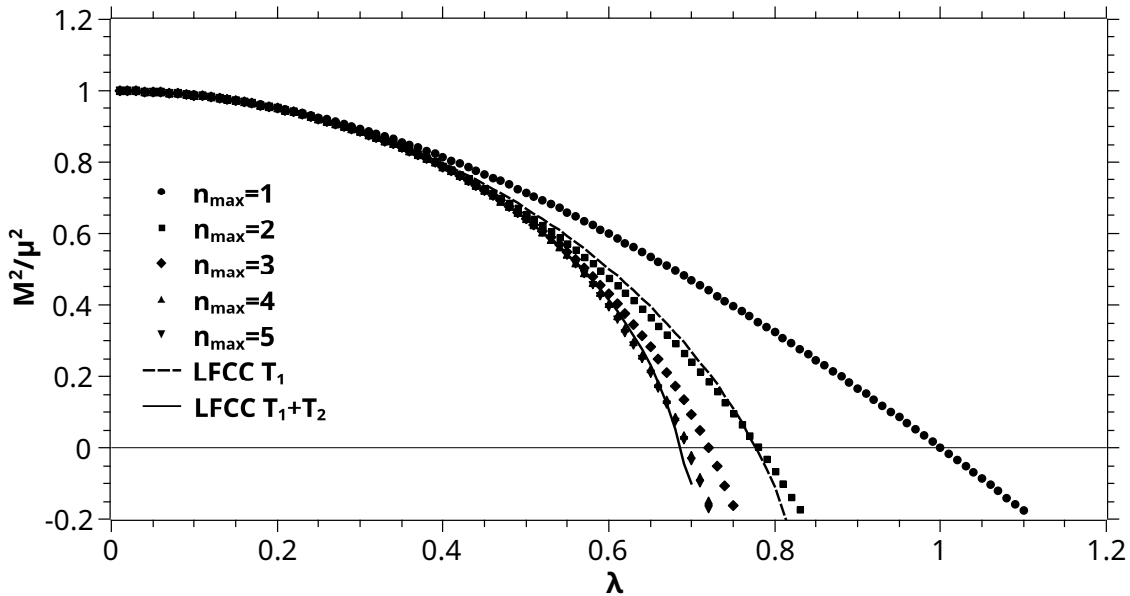}}
\caption{\label{fig:t2mt1}
Same as Fig.~\protect\ref{fig:t1mt01} but for a constituent mass ratio of 
$\tilde{m}=1$ and with both LFCC approximations $T=T_1$ and $T=T_1+T_2$.
The basis sets in each Fock sector were limited to orders $N=2$, 6, 12, 10, 
and 8 for $n_{\rm max}=1$, 2, 3, 4, and 5, respectively.  The basis
sets for the LFCC results have maximum orders of $N_1=5$ and $N_2=5$.
}
\end{figure}

\begin{figure}[ht]
\vspace{0.2in}
\centerline{\includegraphics[width=15cm]{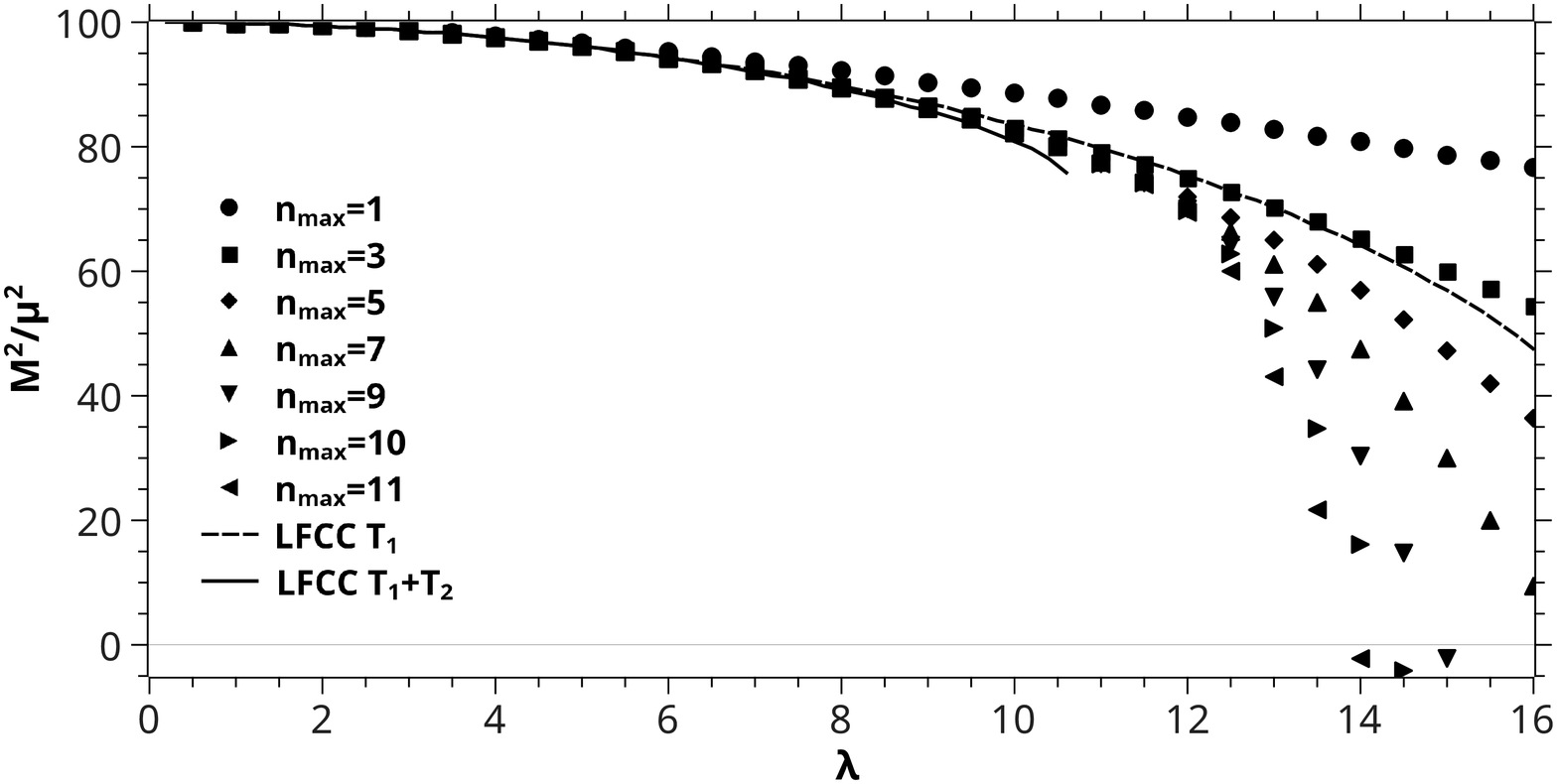}}
\caption{\label{fig:t2mt10}
Same as Fig.~\protect\ref{fig:t2mt1} but for the mass
ratio of the constituents is $\tilde{m}=10$.   The basis sets
in each Fock sector were limited to orders $N=5$, 5, 4, 3, and 3 for $n_{\rm max}=1$,
2, 3, 4, and 5, respectively, and to $N=2$ for all higher Fock sectors. The basis
sets for the LFCC results have maximum orders of $N_1=5$ and $N_2=3$. In this case,
the Fock-space expansion has not converged near $M=0$. Also,
the nonlinear system solver failed to converge for the LFCC approximation
with $\lambda$ beyond 10.6 when $T_2$ was included.
}
\end{figure}

The results for the mass $M$ in the Fock-state expansion method
are shown in Figs.~\ref{fig:t1mt01}-\ref{fig:t2mt10}.  Both the basis
size and the Fock-space limit are increased to achieve convergence for
the lowest eigenstate; however, for the ultrarelativistic case of $\tilde{m}=m/\mu=10$,
convergence of the Fock-space expansion is not achieved for stronger coupling values,
as can be seen in Fig.~\ref{fig:t2mt10}.  On the other hand, convergence for
the nonrelativistic case of $\tilde{m}=0.1$ is almost immediate.

From the solutions to the LFCC equations, we compute the mass eigenvalues $M$
and the relative probabilities of the one and two-neutral Fock sectors. 
The masses are shown in Figs.~\ref{fig:t1mt01}-\ref{fig:t2mt10}, where we plot
results for both $T_1$ alone and $T_1+T_2$.  

\begin{figure}[ht]
\vspace{0.2in}
\centerline{\includegraphics[width=15cm]{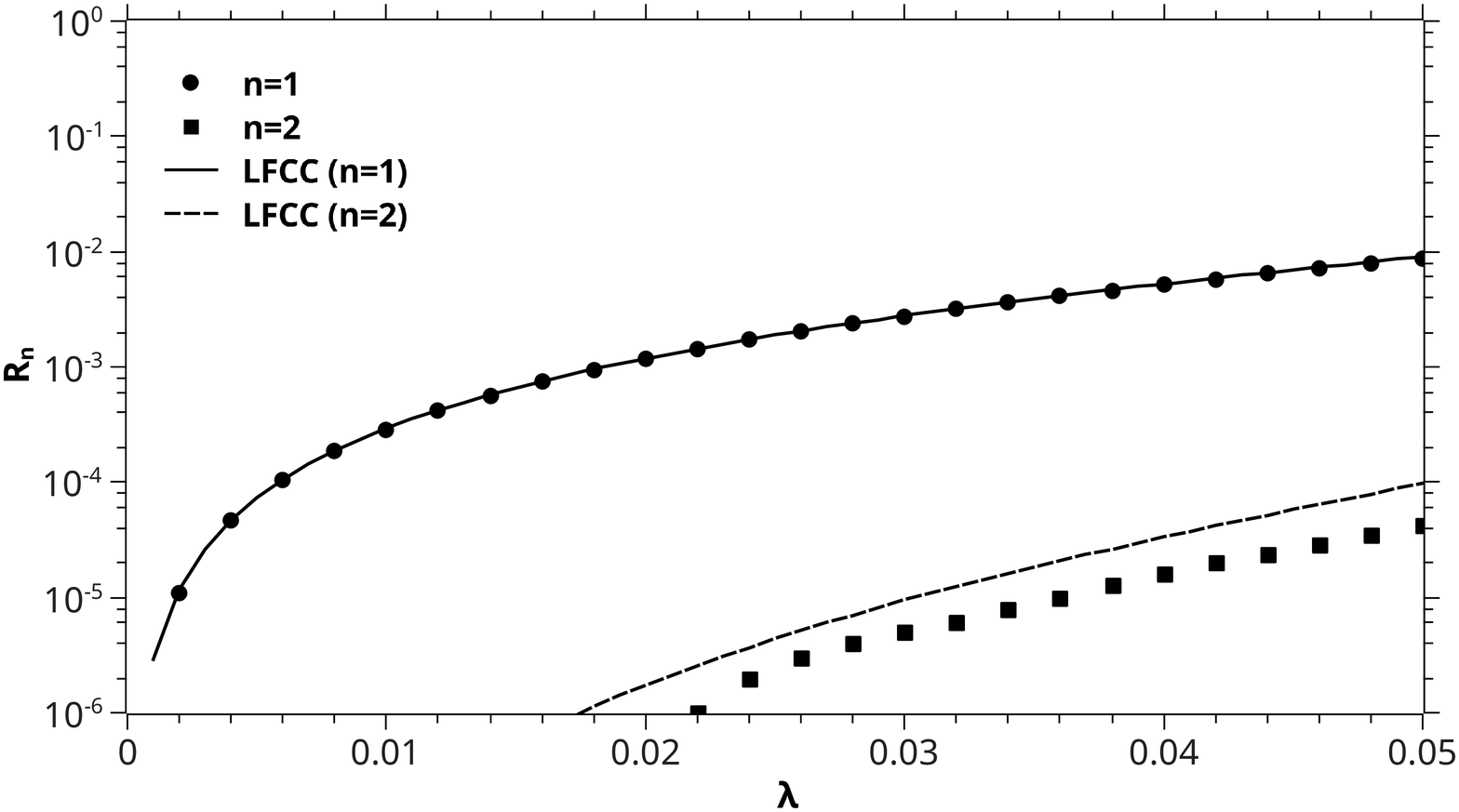}}
\caption{\label{fig:mt01RelProb}
Relative probabilities $R_n$ for a sequence of Fock sectors as functions
of the dimensionless coupling $\lambda$ for a constituent mass ratio of
$\tilde{m}=0.1$.  Results for the one and two-neutrals Fock sectors in
the LFCC approximation are also included.
}
\end{figure}

\begin{figure}[ht]
\vspace{0.2in}
\centerline{\includegraphics[width=15cm]{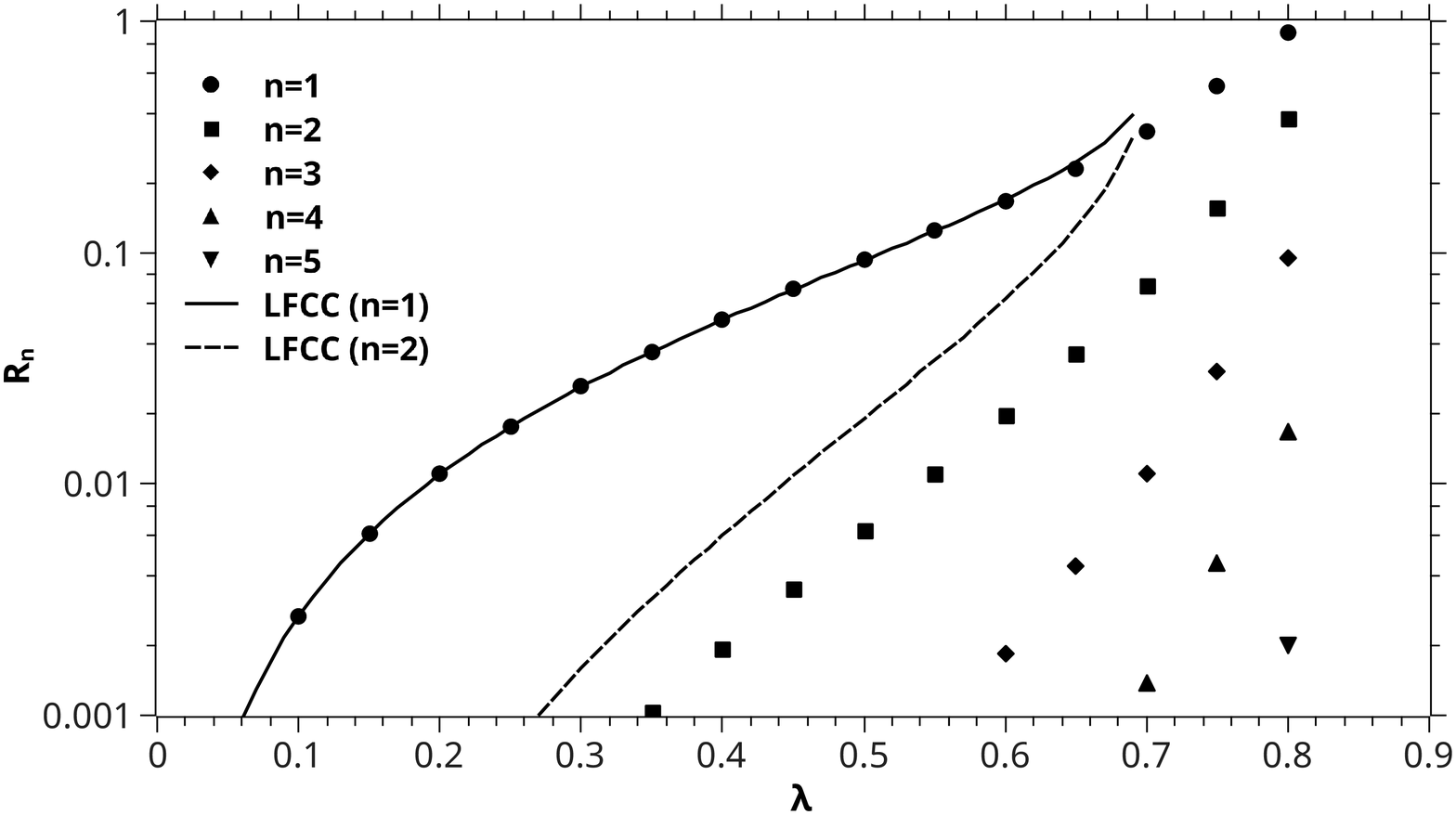}}
\caption{\label{fig:mt1RelProb}
Same as Fig.~\protect\ref{fig:mt01RelProb} but for $\tilde{m}=1$.
}
\end{figure}

\begin{figure}[ht]
\vspace{0.2in}
\centerline{\includegraphics[width=15cm]{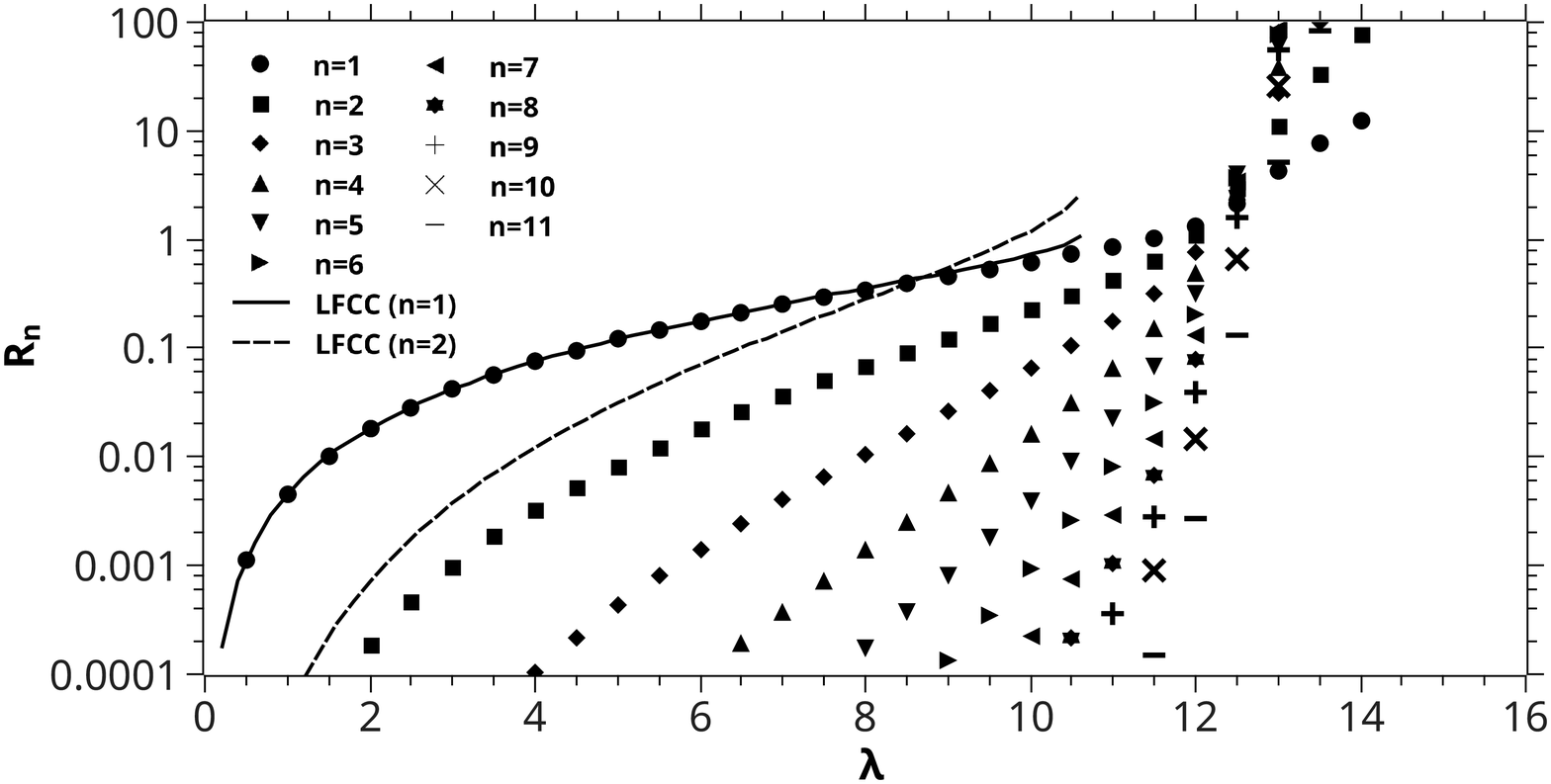}}
\caption{\label{fig:mt10RelProb}
Same as Fig.~\protect\ref{fig:mt01RelProb} but for $\tilde{m}=10$.
}
\end{figure}

Results for relative probabilities are plotted in Figs.~\ref{fig:mt01RelProb}-\ref{fig:mt10RelProb}.
These show that as the neutral constituents become lighter, making $\tilde{m}$
larger, the importance of the higher Fock sectors increases dramatically.
The LFCC approximation for the one-neutral Fock wave function yields a nearly
exact match to the one-neutral relative probability; this is seen in 
Figs.~\ref{fig:mt01RelProb}-\ref{fig:mt10RelProb}, where the solid line 
representing the LFCC result passes through the points from the
converged Fock-state-expansion results for the one-neutral probabilities.
We interpret this agreement to mean that the effect of the higher Fock sectors on
the one-neutral wave function is well represented by the LFCC approximation
to these higher sectors.


The results show that the LFCC truncation to $T_1+T_2$ is sufficient
to replicate the converged Fock-state expansion results, with $T_1$
alone just as good as a two or three-neutral Fock-sector truncation.
Thus the LFCC approximation, using only the two functions $t_1(y)$
and $t_2(y_1,y_2)$ of one and two variables, respectively, is
sufficient to represent information that the Fock-space expansion
encodes in many more wave functions.  In addition, the number of
basis functions required to represent the Fock wave functions
is significantly greater than the number required for the LFCC
functions.  Thus, the matrix representation is much smaller for
the LFCC approximation, which is ample compensation for its
nonlinearity.

The failure of the nonlinear solver to converge\footnote{A calculation
done using {\em Mathematica} also fails to converge and instead
indicates that the desired physical solution has ceased to exist.} 
for strong coupling in the ultrarelativistic case occurs in the same
coupling range where the Fock-state expansion fails to
converge.  This is near where $M$ tends to zero and may
be indicative of the incompleteness of theory.  Quenching
may have stabilized the spectrum, but the theory is no
longer a complete quantum theory.  As discussed in the
Introduction, a similar lack of solution convergence has been observed
in $\phi^4$ theory~\cite{LFCCphi4}.

\section{Summary} \label{sec:summary}

We have shown that the LFCC approximation provides an efficient
representation of a massive eigenstate in quenched scalar Yukawa 
theory.  We have also found that the LFCC approximation converges
quickly as more terms are added to the $T$ operator.  From
a numerical standpoint, there is also an efficiency in the
basis size required for a matrix representation of the 
fundamental equations; the LFCC functions are fewer in number
than the Fock wave functions, depend on fewer variables, and
need fewer basis functions for their accurate representation.

In doing these calculations, we have developed diagrammatic
methods for the construction of the LFCC equations.  These
significantly reduce the effort involved, compared to 
literally carrying out contractions of creation
and annihilation operators in matrix elements of the
effective LFCC Hamiltonian.  Extension to other theories
should be straightforward.

\acknowledgments
This work was supported in part by 
the Minnesota Supercomputing Institute through
grants of computing time and benefited from 
participation in the workshop on
Hamiltonian methods in strongly coupled quantum field theory
supported by the Simons Collaboration on the Nonperturbative Bootstrap.
The operator diagrams were drawn with JaxoDraw~\cite{JaxoDraw}.

\appendix

\section{Numerical methods}  \label{sec:methods}

\subsection{Fock-state expansion}

We solve the coupled system (\ref{eq:FSsystem}) for the Fock-state wave functions
$\psi_n$ in (\ref{eq:FSexpansion}) by first expanding the wave functions in a
simple polynomial basis
\be
\psi_n(y_1,\ldots,y_n)=\sqrt{y_1\cdots y_n(1-\sum_i y_i)}
       \sum_{mj}^N c_{mj}^{(n)}P_{mj}^{(n)}(y_1,\ldots,y_n),
\ee
where $m$ is the order of the polynomial $P_{mj}^{(n)}$,
$j$ is an index that differentiates distinct polynomials of
the same order (which is nontrivial for multivariate polynomials),
$N$ is the maximum order included,
and the $c_{mi}^{(n)}$ are unknown coefficients to be obtained.  The 
polynomials are chosen to be simple monomials, suitably symmetrized
but not orthogonal.  They take the form
\be
P_{mj}^{(n)}(y_1,\ldots,y_n)=y_1^{j_1}y_2^{j_2}\ldots y_n^{j_n}+\cdots,
\ee
with $\sum_i^n j_i=m$.  The truncation of the basis to the order $N$
is, of course, an approximation necessary for a finite matrix
representation; we study convergence with the respect to this
truncation, allowing $N$ to be different for each Fock sector.

Projection of the nth equation onto each basis function,
$\sqrt{y_1\cdots y_n(1-\sum_i y_i)}P_{m'j'}^{(n)}(y_1,\ldots,y_n)$,
yields a matrix representation of the original coupled system
\be
\sum_{mj}\left[T^{(n)}_{m'j',mj}c_{mj}^{(n)}
    +V^{(n,n+1)}_{m'j',mj}c_{mj}^{(n+1)}+V^{(n,n-1)}_{m'j',mj}c_{mj}^{(n-1)}\right]
    =\frac{M^2}{\mu^2}\sum_{mj}S^{(n)}_{m'j',mj}c_{mj}^{(n)}.
\ee
The individual matrices are
\be
T^{(n)}_{m'j',mj}=\int \prod_i^n dy_i\left[\tilde{m}^2y_1\cdots y_n
                 +ny_2\cdots y_n\right] P_{m'j'}^{(n)}(y_1,\ldots,y_n)P_{mj}^{(n)}(y_1,\ldots,y_n),
\ee
\be
V^{(n,n+1)}_{m'j',mj}=\lambda\sqrt{n+1} \int \prod_i^n dy_i\int_0^{1-\sum_i y_i}dx\, y_1\cdots y_n
                  P_{m'j'}^{(n)}(y_1,\ldots,y_n)P_{mj}^{(n+1)}(y_1,\ldots,y_n,x),
\ee
\be
V^{(n,n-1)}_{m'j',mj}=\lambda\sqrt{n} \int \prod_i^n dy_i\, y_1\cdots y_{n-1}
                  P_{m'j'}^{(n)}(y_1,\ldots,y_n)P_{mj}^{(n-1)}(y_1,\ldots,y_{n-1}),
\ee
and
\be
S^{(n)}_{m'j',mj}=\int \prod_i^n dy_i\, y_1\cdots y_n (1-\sum_i y_i) 
                     P_{m'j'}^{(n)}(y_1,\ldots,y_n)P_{mj}^{(n)}(y_1,\ldots,y_n).
\ee
The integrals can be done analytically in terms of the generalized $\beta$ function
\be  \label{eq:betafn}
\int dx_1 \cdots dx_n \, x_1^{k_1}\cdots x_n^{k_n}(1-x_1-\cdots-x_n)=\frac{k_1!\cdots k_n!}{(k_1+\cdots+k_n+n+2)!}
\ee
This allows for efficient calculation of all the integrals, with the different $\beta$-function
evaluations done recursively and stored for use.

If the basis functions were orthogonal, $S^{(n)}$ would be diagonal, of course.  However,
we implicitly orthogonalize the basis by performing a singular-value decomposition
$S^{(n)}=U^{(n)}W^{(n)}U^{(n)T}$.  The columns of the matrix $U^{(n)}$
are the eigenvectors of $S^{(n)}$, and $W^{(n)}$ is a diagonal
matrix of the eigenvalues.  The $U$ matrices then define an orthogonal
transformation to new vectors of coefficients $\vec c^{\,(n)\prime}=(W^{(n)})^{1/2}U^{(n)T}\vec c^{\,(n)}$
and new matrices, such as $T^{(n)\prime}=(W^{(n)})^{-1/2}U^{(n)T} T^{(n)} U^{(n)}(W^{(n)})^{-1/2}$.
The new matrix problem is no longer of the generalized type, but simply
\be
\sum_{mj}\left[T^{(n)\prime}_{m'j',mj}c_{mj}^{(n)\prime}
    +V^{(n,n+1)\prime}_{m'j',mj}c_{mj}^{(n+1)\prime}+V^{(n,n-1)\prime}_{m'j',mj}c_{mj}^{(n-1)\prime}\right]
    =\frac{M^2}{\mu^2}c_{m'j'}^{(n)\prime}.
\ee
The lowest eigenvalue is extracted by standard procedures for symmetric matrices.

\begin{figure}[ht]
\vspace{0.2in}
\centerline{\includegraphics[width=15cm]{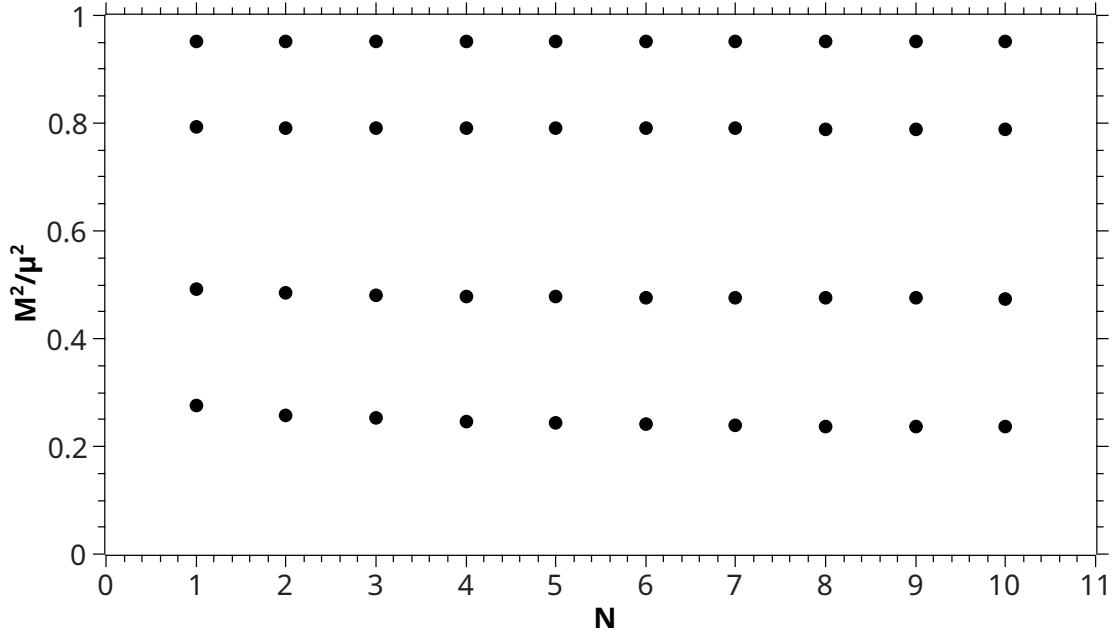}}
\caption{\label{fig:mt1FSbasisConv}
The mass eigenvalue as a function of the basis order $N$ in the top
Fock sector of two neutrals for selected
coupling strengths $\lambda=0.2$, 0.4, 0.6, and 0.7.  The mass values
decrease with increasing $\lambda$.  The constituent mass ratio $\tilde{m}$ is 
equal to 1.  The maximum polynomial order in the one-neutral sector is 2.
}
\end{figure}

The convergence of such a calculation, with respect to the basis size, is  illustrated
in Fig.~\ref{fig:mt1FSbasisConv}.  Convergence is quite rapid in general; for stronger
coupling values, near where $M$ becomes zero, larger basis sizes are needed.

\subsection{LFCC approximation}

To solve the LFCC equations for $t_1$ and $t_2$, given in (\ref{eq:t1final})
and (\ref{eq:t2}), we expand these functions in the basis set used 
for the Fock-state wave functions as
\be
t_1(y)=\sqrt{y(1-y)}\sum_m^{N_1} a_m P_m^{(1)}(y), \;\;
t_2(y_1,y_2)=\sqrt{y_1 y_2 (1-y_1-y_2)}\sum_m^{N_2}b_m P_m^{(2)}(y_1,y_2).
\ee
Here the index $m$ represents both the order and implicitly, in
the case of two variables, the
distinction between linearly independent polynomials of the 
same order.  The equation for $t_1$ is projected onto
the single-variable basis functions $\sqrt{y(1-y)}P_{m'}^{(1)}(y)$,
and the equation for $t_2$ is projected onto
$\sqrt{y_1 y_2(1-y_1-y_2)}P_{m'}^{(2)}(y_1,y_2)$.
The matrix representation of the equation for $t_1$ is then
\bea
0&=&(\tilde{m}^2+\frac{\lambda}{2}\Delta)A_{m'm}^1a_m-(\tilde{m}^2+\lambda\Delta)A_{m'm}^2a_m
    +B_{m'm}a_m+\lambda C_{m'}+\frac{\lambda}{2} D_{m'ml}a_m a_l
    +2\lambda F_{m'm}b_m  \nonumber \\
\eea
and that for $t_2$ is
\bea
0&=&2\left[(\tilde{m}^2+\frac{\lambda}{2}\Delta)G_{m'm}^1 
     -(\tilde{m}^2+\lambda\Delta) G_{m'm}^2 
     +G_{m'm}^3 \right]b_m \\
&&     +\left[(\tilde{m}^2+\frac{\lambda}{3}\Delta)H_{m'ml}^1
     -2(\tilde{m}^2+\frac{\lambda}{2}\Delta)H_{m'ml}^2
     +(\tilde{m}^2+\lambda\Delta)H_{m'ml}^3 
     +H_{m'ml}^4\right] a_m a_l
\nonumber \\
&&     +2\lambda I_{m'm} a_m
     +\frac{\lambda}{3}J_{m'mkl} a_m a_k a_l
     +\lambda K_{m'ml} b_m a_l,
\nonumber
\eea
with sums over repeated indices implied,
\be
\Delta=\sum_m C_m a_m
\ee
and the associated matrices defined by
\bea
A_{m'm}^1&=&\int_0^1 dy\, y P_{m'}^{(1)}(y) P_m^{(1)}(y), \\
A_{m'm}^2&=&\int_0^1 dy\, y (1-y) P_{m'}^{(1)}(y) P_m^{(1)}(y) \\
B_{m'm}&=&\int_0^1 dy\, (1-y) P_{m'}^{(1)}(y) P_m^{(1)}(y) \\
C_{m'}&=&\int_0^1 dy P_{m'}^{(1)}(y) \\
D_{m'ml}&=&\int_0^1 dy\int_0^{1-y} dx\frac{y}{1-x}P_{m'}^{(1)}(y)P_m^{(1)}(\frac{y}{1-x})P_l^{(1)}(x) \\
        &=&\int_0^1 dx \int_0^1 dz\, z(1-x) P_{m'}^{(1)}(z(1-x))P_m^{(1)}(z)P_l^{(1)}(x) \nonumber \\
F_{m'm}&=&\int_0^1 dy \int_0^{1-y} dx\, y P_{m'}^{(1)}(y)P_m^{(2)}(y,x) \\
G_{m'm}^1&=&\int_0^1 dy_1 dy_2\, y_1 y_2 P_{m'}^{(2)}(y_1,y_2)P_m^{(2)}(y_1,y_2) \\
G_{m'm}^2&=&\int_0^1 dy_1 dy_2\, y_1 y_2 (1-y_1-y_2) P_{m'}^{(2)}(y_1,y_2)P_m^{(2)}(y_1,y_2) \\
G_{m'm}^3&=&2\int_0^1 dy_1 dy_2\, y_2 (1-y_1-y_2) P_{m'}^{(2)}(y_1,y_2)P_m^{(2)}(y_1,y_2)  \\
H_{m'ml}^1&=&\int_0^1 dy_1 dz_2\, y_1 z_2 (1-y_1) P_{m'}^{(2)}(y_1,z_2(1-y_1))P_m^{(1)}(y_1)P_l^{(1)}(z_2) \\
H_{m'ml}^2&=&\int_0^1 dy_1 dz_2\, y_1 z_2(1-y_1)(1-z_2) P_{m'}^{(2)}(y_1,z_2(1-y_1))P_m^{(1)}(y_1)P_l^{(1)}(z_2) \\
H_{m'ml}^3&=&\int_0^1 dy_1 dz_2\, y_1 z_2(1-y_1)^2(1-z_2) P_{m'}^{(2)}(y_1,z_2(1-y_1))P_m^{(1)}(y_1)P_l^{(1)}(z_2) \\
H_{m'ml}^4&=&\int_0^1 dy_1 dz_2\, y_1 (1-y_1)(1-z_2) P_{m'}^{(2)}(y_1,z_2(1-y_1))P_m^{(1)}(y_1)P_l^{(1)}(z_2) \\
     &&       -\int_0^1 dy_1 dz_2\,  z_2(1-y_1)^2(1-z_2) P_{m'}^{(2)}(y_1,z_2(1-y_1))P_m^{(1)}(y_1)P_l^{(1)}(z_2)
            \nonumber \\
I_{m'm}&=&\int_0^1 dy_1 dy_2\, y_1 P_{m'}^{(2)}(y_1,y_2) P_m^{(1)}(y_1) \\
     &&   -\int_0^1 dy_1 dz_2\, z_2(1-y_1)(1-z_2)P_{m'}^{(2)}(y_1,z_2(1-y_1)) P_m^{(1)}(z_2) \nonumber \\
J_{m'mkl}&=&\int_0^1 dy_1 dz_1 dz_2\left[ y_1 z_2 (1-y_1)(1-z_1) P_{m'}^{(2)}(y_1,z_2(1-y_1)(1-z_1)) \right. \\
       &&  -3 z_1 (1-y_1)z_2(1-z_1(1-y_1))^2(1-z_2) P_{m'}^{(2)}(z_1(1-y_1),z_2(1-y_1)(1-z_1))
       \nonumber \\
       &&  \left.+z_1 (1-y_1)z_2(1-z_1)(1-y_1)) P_{m'}^{(2)}(z_1(1-y_1),z_2(1-y_1)(1-z_1)) \right] 
       \nonumber \\
       &&   \rule{0.5in}{0mm}  \times  P_m^{(1)}(z_2)P_k^{(1)}(z_1)P_l^{(1)}(y_1)  \nonumber \\
K_{m'ml}&=&\int_0^1 dy_1 dz_1 dz_2 \left[z_1 z_2 (1-y_1)^2(1-z_1)^2 P_{m'}^{(2)}(z_1(1-y_1),z_2(1-y_1)(1-z_1)) \right. \\
     &&  \rule{3in}{0mm}    \times    P_m^{(2)}(z_1,z_2(1-z_1)) P_l^{(1)}(y_1)  \nonumber \\
     &&       +2 y_1 z_2 (1-y_1)(1-z_2) P_{m'}^{(2)}(y_1,z_2(1-y_1))
                                    P_m^{(2)}(z_1(1-z_2),z_2)) P_l^{(1)}(y_1) \nonumber \\
     &&       -4 y_1 z_2 (1-y_1)^2 (1-z_2) P_{m'}^{(2)}(y_1,z_2(1-y_1))
                                    P_m^{(2)}(z_1(1-y_1),y_1)) P_l^{(1)}(z_2) \nonumber \\
     &&    \left.   +2 y_1 z_2 (1-y_1)^2(1-z_1) P_{m'}^{(2)}(y_1,z_2(1-y_1)(1-z_1))
                                    P_m^{(2)}(y_1,z_1(1-y_1)) P_l^{(1)}(z_2)\right]. \nonumber
\eea
For the $D$ matrix, a change of variables has been shown explicitly; similar rescalings are done
for many of the other matrices.  These rescalings arrange for the arguments of the polynomials
to be polynomials and for all integration ranges to be from 0 to 1.  The integrals are then
linear combinations of simple integrals of monomials.

The nonlinear matrix equations obtained in this way are then solved 
by a modification of the Powell hybrid method~\cite{Powell} as implemented
in the general nonlinear equation solver `hybrj' of the MINPACK set of subroutines~\cite{minpack}.
The method is recursive; the initial guess for the unknown coefficients is taken to be zero for
the lowest coupling strength and, as an increasing series of coupling strengths is
considered, the next initial guess is the solution for the previous coupling strength.

\begin{figure}[ht]
\vspace{0.2in}
\centerline{\includegraphics[width=15cm]{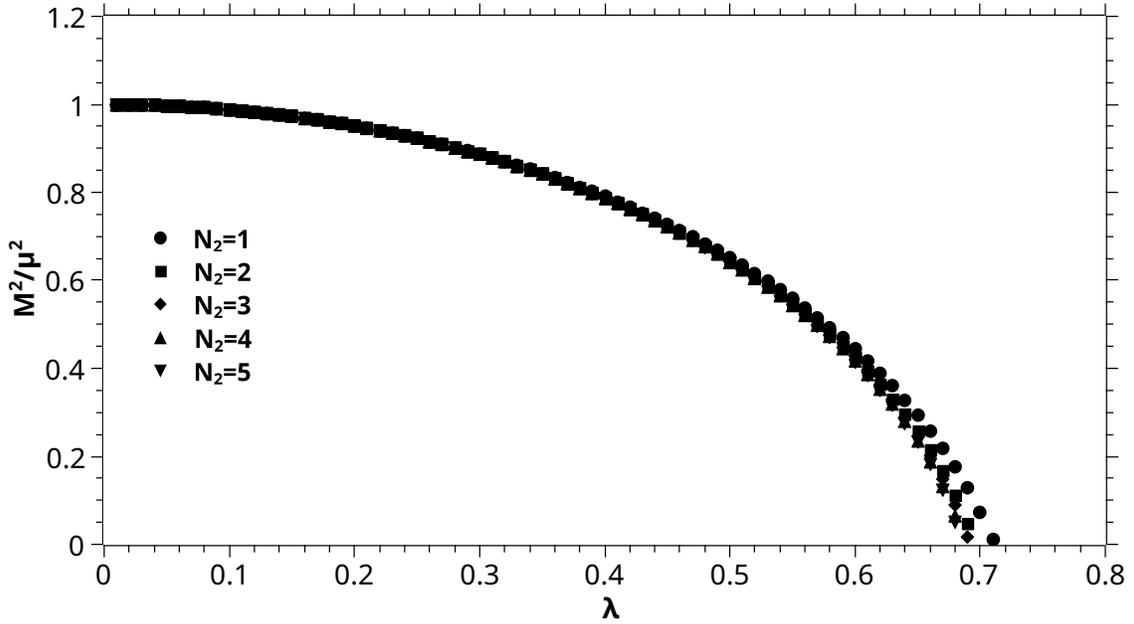}}
\caption{\label{fig:t2mt1Conv}
LFCC results for the mass eigenvalue ratio $M^2/\mu^2$ as a function of the dimensionless
coupling $\lambda$ for a range of basis sizes for $t_2$.  The basis set
was limited to maximum order of $N_2=1$ through 5, with the $t_1$
basis size set at maximum order $N_1=2$. The mass
ratio of the constituents is $\tilde{m}\equiv m/\mu=1$. 
}
\end{figure}

For the case where only $T_1$ is included and we solve only (\ref{eq:t1final}) for $t_1$ with
$t_2=0$, convergence with respect to basis size is very rapid when $\tilde{m}=1$.  The results for $N_1=1$
and $N_1=2$ are indistinguishable on a graph.  For the full solution, with both $T_1$ and $T_2$
included, the dependence on $N_2$, the maximum order for the $t_2$ basis, is shown in Fig.~\ref{fig:t2mt1Conv}.
Convergence is again quite rapid, except for stronger coupling where $M^2$ approaches zero.  For smaller
and larger values of $\tilde{m}$, convergence is slower for $t_1$, requiring $N_1=9$ for
$\tilde{m}=0.1$ and $N_1=5$ for $\tilde{m}=10$.  Convergence for $t_2$ is quicker,
using $N_2=3$, except for strong coupling in the case of $\tilde{m}=10$ where the
nonlinear equation solver was unable to converge to a solution.

\section{Rules for diagrams} \label{sec:rules}

\begin{figure}[ht]
\vspace{0.2in}
\centerline{\includegraphics[width=15cm]{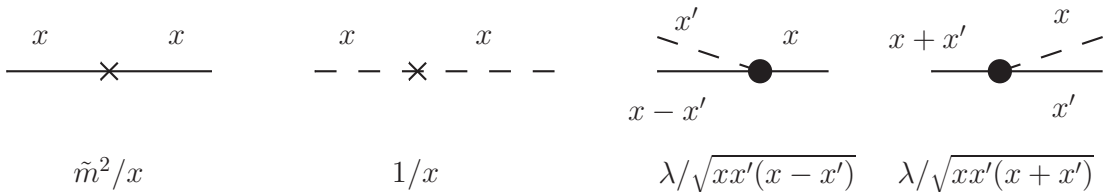}}
\caption{\label{fig:Pminus}
Diagrammatic representation of the terms in $\Pminus$
and their corresponding expressions.  Solid lines represent
the charged scalar and dashed, the neutral.  A cross designates
a mass term.  The diagrams represent operators acting to the
right; for example, the last diagram corresponds to the 
annihilation of a neutral.
}
\end{figure}

\begin{figure}[ht]
\vspace{0.2in}
\centerline{\includegraphics[width=8cm]{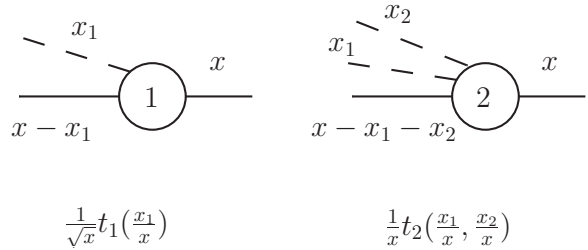}}
\caption{\label{fig:T1T2}
Diagrammatic representation of the $T_1$ and $T_2$ operators,
and their corresponding expressions, acting to the right
and creating one or two neutrals, respectively, by first
annihilating a charged scalar with momentum fraction $x$.
}
\end{figure}

Although the LFCC equations for the $t$ functions
can be constructed by carrying out the contractions of
creation and annihilation operators, the construction can
be simplified by use of a set of rules for operator diagrams
that depict the structure of the contractions.  The rules
are as follows:

\begin{enumerate}
 \item Represent the terms of $\Pminus$ by crosses for the charged
 and neutral mass terms and simple vertices for neutral creation and
 annihilation, as shown in Fig.~\ref{fig:Pminus}.
 \item Represent $T_1$ and $T_2$ by the vertices shown in Fig.~\ref{fig:T1T2}.
 \item For each Fock-sector projection, draw all possible diagrams
 connecting the valence state to that Fock sector.  The connections
 between vertices and/or crosses represent contractions. Each
 diagram must include a term from  $\Pminus$ once and only once
 and may include as many $T_1$ and/or $T_2$ vertices as needed,
 to the left and right of the $\Pminus$ insertion, to reach the
 chosen sector.
 \item In each diagram, label each line with a momentum fraction,
 starting from 1 for the line acting on the charge-one valence
 state on the right and ending with $y_1$ through $y_n$ for the 
 $n$ neutrals in the projected sector on the left;
 conserve momentum at each vertex.
 \item Construct the expression corresponding to the
 diagram from the individual vertices and crosses, and
 integrate over any loop momentum fractions, with the upper limit
 set by the fractions entering and leaving the loop.
 \item For each product of $m$ $T_1$ and $T_2$ vertices to
 the left, include a factor of $(-1)^m/m!$ and for each to
 right, a factor of $1/m!$; these come from the expansion
 of the exponential $e^{\pm(T_1+T_2)}$.
 \item Symmetrize with respect to permutations of $y_1,\ldots,y_n$
 and with respect to the neutral lines from $T_2$ vertices.
\end{enumerate}

As an almost trivial example, the diagrams contributing to the
terms on the right of the valence equation (\ref{eq:M})
are shown in Fig.~\ref{fig:valence}.  A less trivial example
is the set of diagrams for the one-neutral projection,
shown in Fig.~\ref{fig:oneneutral}.  Except for the $\frac12\Pint T_1^2$
term in (\ref{eq:oneneutral}), there is only one diagram for each term in
$\ob{\Pminus}$; for $\frac12\Pint T_1^2$ there are two.
The rules then yield (\ref{eq:t1}).


\end{document}